\documentclass{nature}
\usepackage{setspace}
\usepackage{amsfonts}
\usepackage{amssymb}
\usepackage{amsmath}

\usepackage{float}
\usepackage{placeins}
\usepackage{graphicx}

\usepackage[export]{adjustbox}

\usepackage[bf,bf,bf]{caption}

\renewcommand{\figurename}{Figure}
\renewcommand{\tablename}{Table}

\usepackage[super,comma,sort&compress]{natbib}

\usepackage{hyperref}
\usepackage{verbatim}

\usepackage{ulem} %provides various types of underlining that can stretch between words and be broken across lines
\usepackage{rotating} % rotate any object by arbitrary angle
\usepackage{soul}%h y p h e n -a t a b l e l e t t e r s p a c i n g ( s p a c i n g o u t ) , underlining and some derivatives such as overstriking and highlighting

\usepackage{color} %change color of text

\usepackage[colorinlistoftodos]{todonotes}

\makeatletter
\newsavebox\myboxA
\newsavebox\myboxB
\newlength\mylenA

\newcommand*\xoverline[2][0.75]{%
    \sbox{\myboxA}{$\m@th#2$}%
    \setbox\myboxB\null% Phantom box
    \ht\myboxB=\ht\myboxA%
    \dp\myboxB=\dp\myboxA%
    \wd\myboxB=#1\wd\myboxA% Scale phantom
    \sbox\myboxB{$\m@th\overline{\copy\myboxB}$}%  Overlined phantom
    \setlength\mylenA{\the\wd\myboxA}%   calc width diff
    \addtolength\mylenA{-\the\wd\myboxB}%
    \ifdim\wd\myboxB<\wd\myboxA%
       \rlap{\hskip 0.5\mylenA\usebox\myboxB}{\usebox\myboxA}%
    \else
        \hskip -0.5\mylenA\rlap{\usebox\myboxA}{\hskip 0.5\mylenA\usebox\myboxB}%
    \fi}
\makeatother

\usepackage{lscape}
\usepackage{mathtools}
\usepackage{gensymb}
\usepackage{hyperref}
\usepackage{subfigure}
\hypersetup{hidelinks}
\usepackage{graphicx}
\usepackage{textcomp, gensymb}
\usepackage{soul,xcolor}

\newcommand{\panel}[1]{\textbf{#1}}
\newcommand{\Mpc}[1]{\ensuremath{~\rm{Mpc}}}
\newcommand{\kms}[1]{\ensuremath{~\rm{km/s}}}
\begin{document}
\captionsetup[figure]{name={Fig.}}

\title{A younger Universe implied by satellite pair correlations from SDSS observations of massive galaxy groups
}%{Velocity correlations of satellite pairs around massive galaxies}

%% Notice placement of commas and superscripts and use of &
%% in the author list

%\author{Qing Gu}
\author{
Qing Gu$^{1,2,3}$; 
Qi Guo$^{*1,2,3}$;
Marius Cautun$^{4}$;
Shi Shao$^{*1}$;
Wenxiang Pei$^{1,2,3}$;
Wenting Wang$^{5,6}$;
Liang Gao$^{1,2,3,7}$;
Jie Wang$^{1,2,3}$
}

\maketitle

\begin{affiliations}
\item Key Laboratory for Computational Astrophysics, National Astronomical Observatories, Chinese Academy of Sciences, Beijing 100101, China;
\item Institute for Frontiers in Astronomy and Astrophysics, Beijing Normal University, Beijing 102206, China
\item School of Astronomy and Space Science, University of Chinese Academy of Sciences, Beijing 10049, China;
\item Leiden Observatory, Leiden University, PO Box 9513, NL-2300 RA Leiden, the Netherlands;
\item Department of Astronomy, Shanghai Jiao Tong University, Shanghai 200240, China;
\item Shanghai Key Laboratory for Particle Physics and Cosmology, Shanghai 200240, China
\item School of Physics and Microelectronics, Zhengzhou University, Zhengzhou, 450001, China
\end{affiliations}

\begin{abstract}

Many of the satellites of galactic-mass systems such as the Miky Way, Andromeda and Centaurus A show evidence of coherent motions to a larger extent than most of the systems predicted by the standard cosmological model. It is an open question if correlations in satellite orbits are present in systems of different masses. Here , we report an analysis of the kinematics of satellite galaxies around massive galaxy groups. Unlike what is seen in Milky Way analogues, we find an excess of diametrically opposed pairs of satellites that have line-of-sight velocity offsets from the central galaxy of the same sign. 
This corresponds to a $\pmb{6.0\sigma}$ ($\pmb{p}$-value $\pmb{=\ 9.9\times10^{-10}}$) detection of non-random satellite motions. Such excess is predicted by up-to-date cosmological simulations but the magnitude of the effect is considerably lower than in observations. The observational data is discrepant at the $\pmb{4.1\sigma}$ and $\pmb{3.6\sigma}$ level with the expectations of the Millennium and the Illustris TNG300 cosmological simulations, potentially indicating that massive galaxy groups assembled later in the real Universe. The detection of velocity correlations of satellite galaxies and tension with theoretical predictions is robust against changes in sample selection. Using the largest sample to date, our findings demonstrate that the motions of satellite galaxies represent a challenge to the current cosmological model.
\end{abstract}

Galaxies and their extended dark matter halos assemble hierarchically within the standard Lambda-Cold-Dark-Matter ($\Lambda$CDM) cosmological model, that is through the merger of many smaller objects. A key probe of this process is the motion of satellite galaxies, which encodes information about the structure and growth of host halos\cite{Frenk2012,Shao2021}. This is why for decades a lot of attention has been paid to study the motion of satellites around nearby central galaxies. One prime example is our Galaxy, whose satellites currently reside in a planar structure\cite{Lynden-Bell1976} and most of which also orbit in this common plane\cite{Metz2009,Pawlowski2016,Pawlowski2020}. Co-rotating planar structures are also found in subsamples of satellites around the Andromeda (M31)\cite{Metz2007,Conn2012, Ibata2013} and Centaurus A\cite{Muller2021}, which are thought to be a manifestation of the anisotropic accretion of satellite galaxies along cosmic web filaments\cite{Libeskind2005,Buck2015,Shao2018}, group infall of satellite galaxies\cite{Li2008}, galaxy interactions\cite{Hammer2013,Smith2016,Banik2022} or flattening due to the potential field of the large-scale structure \cite{Libeskind2014}. However, the large correlation present in the motion of these subsets of satellites is in the tail of $\Lambda$CDM predictions \cite{Ibata2014R,Cautun2015,Pawlowski2020,Muller2021}, potentially indicating a discrepancy between data and theory. Studies of satellite motion in other galactic-mass systems have proven inconclusive due to the small sample size \cite{Ibata2014N,Cautun2015,Phillips2015,Ibata2015}. However, the small sample size that limited previous studies can be avoided if we investigate more massive systems, such as massive galaxy groups, which is what we report in this article.

\begin{figure*}
\centering 
{
\includegraphics[width=\textwidth]{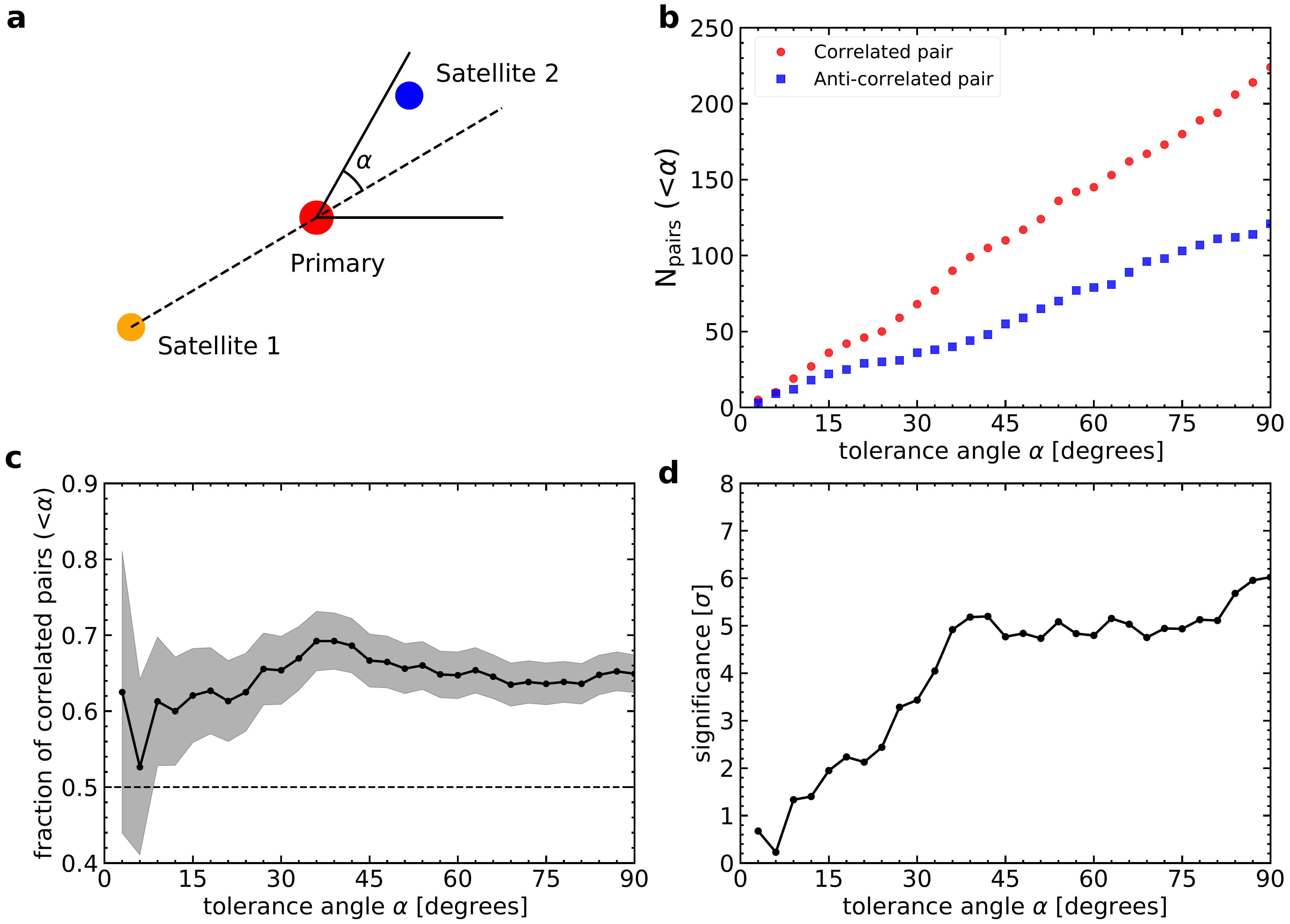}}

\caption{
    {\bf The velocity correlation of the two most massive satellite galaxies in SDSS.} 
    \panel{a)} It illustrates the process of finding a satellite pair on opposite sides of the primary galaxy (big red dot). For each satellite, shown here by the orange dot, we find its counterpart within a tolerance angle, $\alpha$. For our example, the blue dot is a valid counterpart. 
    \panel{b)} The cumulative number of correlated (red circles) and anti-correlated (blue squares) SDSS satellite pairs as a function of tolerance angle. There is a clear excess of correlated pairs for all tolerance angles. 
    \panel{c)} The cumulative fraction of correlated pairs as a function of tolerance angle (black line with dots) compared to the null expectation (dashed line at 0.5) and the 1-$\sigma$ deviation (grey shaded region) expected for bootstrap errors with 1000 samples. The fraction of correlated pairs is $\sim$ 64\%. 
    \panel{d)} The significance of the excess of correlated pairs with respect to a random distribution in units of standard deviations. The significance has a maximum value of 6.0$\sigma$ at $\alpha = 90^{\degree}$, which indicates a robust detection of non-random satellite motions in our sample.
    }
\label{fig:SDSS_top2}
\end{figure*}

We follow the method of quantifying the degree of coherent rotations in satellite systems proposed by Ibata et al\cite{Ibata2014N}. It looks at the velocity correlations between pairs of satellites on opposite sides of their primary (see Fig.\ \ref{fig:SDSS_top2}a). For example, a system with co-rotating satellites observed not exactly face-on will have satellite pairs with anti-correlated line-of-sight velocities. That is, with respect to the primary, one satellite will move towards the observer and the other satellite, on the opposite side, will move away from the observer. The practical implementation of this method is sketched in Fig.\ \ref{fig:SDSS_top2}a. For each satellite, we search for the other satellite on the opposite side of its primary that is found within a certain tolerance angle, $\alpha < 90^{\degree}$. A pair of satellite galaxies is considered correlated if the product of their line-of-sight velocity offsets from the central galaxy is positive, and anti-correlated if it is negative.

We use the seventh data release of the Sloan Digital Sky Survey (SDSS DR7) \cite{Abazajian2009} catalogue with the value-added spectroscopic measurements produced by the group from the Max Planck Institute for Astrophysics and the John Hopkins University (MPA-JHU). Our primary, or central galaxy sample, consists of galaxies with stellar mass between 10$^{11}$ $\rm M_{\odot}$ and 10$^{11.5}$ $\rm M_{\odot}$, a typical stellar mass range for central galaxies in massive galaxy groups or clusters \cite{Guo2009}. They are also brighter than an apparent $r$-band magnitude of $r < 16.6$. We further restrict the primary galaxies selection to low-redshift objects, $z < 0.065$, that satisfy the following isolation criteria: 
(i) they have to be at least one magnitude brighter than any companion within a projected radius, $r_{\rm p} = 0.5$ Mpc, and a line-of-sight velocity difference, $|\Delta v|$ $<$ 1000 km s$^{-1}$, and (ii) they have to be the brightest galaxy within $r_{\rm p}<$ 1 Mpc and $|\Delta v|$ $<$ 1000 km s$^{-1}$. Satellite galaxies are defined as those within a projected distance of 0.1 Mpc $< r_{\rm p} <$ 1 Mpc from the primary and with relative line-of-sight velocity difference, $\sqrt{2} \times 25$ km s$^{-1} <|\Delta v| <$ 1000 km s$^{-1}$. Such selection criteria are often used for the study of galaxy clusters in the literature \cite{Yoon2019,Sohn2021}. The maximum projection distance and line-of-sight velocity difference have been selected to ensure the inclusion of most members of the galaxy clusters. It is worth noting that our sample encompasses a volume that is somewhat larger than the corresponding volume within one virial radius and includes not only satellite galaxies within one virial radius but also galaxies that are falling into the systems, as well as fore/background interlopers (See Methods for further details). As we are unable to distinguish between the three types of galaxies (i.e., genuine satellites, galaxies at infall, and interlopers) in observations, we refer to all galaxies that meet our selection criteria as satellite galaxies. We remove satellites within 0.1 Mpc to avoid deblending mistakes and possible incompleteness due to the fibre collisions \cite{Wang2021} and pose a minimum velocity difference to account for the uncertainties in redshift measurements. We further require that all satellites should be more massive than 10$^{10.08}$ $\rm M_{\odot}$, brighter than $r=17.72$ and with redshifts $z<0.065$. This is to ensure the completeness of satellites based on the survey flux limit (Extended Data Fig.\ \ref{fig:galaxy_selection}). We select only primaries that have at least two satellite galaxies. To ensure a clean sample, we remove primaries and satellites with velocity uncertainties greater than 25 km s$^{-1}$. This leads to a total number of 914 galaxy systems. Finally, we remove those primaries which are outside the survey footprint covered by the New York University Value-Added Galaxy Catalog (NYU-VAGC) spherical polygons, leading to our final sample of 813 galaxy systems.

We focus on the two most massive satellite galaxies in each system. This allows for analysing the largest number of systems and also enables us to investigate fainter primaries. Fig.\ \ref{fig:SDSS_top2}b shows the cumulative number of correlated and anti-correlated satellite pairs as a function of the tolerance angle, $\alpha$. It shows a clear excess of correlated pairs for all values of $\alpha$. To better highlight the difference, in Fig.\ \ref{fig:SDSS_top2}c we show the cumulative fraction of correlated pairs. The fraction at $\alpha=90^{\degree}$ is 64.9\%. The uncertainties are shown for illustration purposes and correspond to the standard deviation of 1000 bootstrap samples. The excess of correlated pairs at the opposite side of their primaries suggests a preference for counter-rotating satellites, opposite to the co-rotating satellites found in Milky Way analogues which have a fraction below 50\%.

To calculate the significance of the excess of correlated pairs, we model the null expectation as a random distribution with a median value of 0.5 for each $\alpha$ value. We account for the ``look-elsewhere effect", i.e. that we did not know {\it a priori} if the correlated pair fraction should be higher or lower than the null expectation of 0.5, and we express the significance in units of standard deviations of a normal distribution. We find a highly significant detection of an excess of correlated satellite pairs that is 6.0$\sigma$ ($p$-value of 9.9$\times10^{-10}$) at $\alpha=90^{\degree}$ (Fig.\ \ref{fig:SDSS_top2}d). 

Central galaxies moving with respect to their host halo or satellite galaxies being misidentified as centrals can both lead to an excess of counter-rotating pairs. However, in order to obtain the observed fraction of correlated satellite pairs, it requires the velocity dispersion of the central galaxies as large as 86\% of the velocity dispersion of the satellite galaxies (Extended Data Fig.\ \ref{fig:toy-model}a), or 56\% of central galaxies being misidentified (Extended Data Fig.\ \ref{fig:toy-model}b). These values are much higher than those found in SDSS\cite{Skibba2011} and in simulations\cite{Ye2017}. When taking into account both effects, the fraction of correlated pairs is 0.59, which is still lower than the value observed in SDSS. (Extended Data Fig.\ \ref{fig:toy-model}c). However, this toy model cannot account for actual contamination from background or foreground systems. It may diminish the signal of excess correlated pairs if we solely include uncorrelated background and foreground systems (Extended Data Fig.\ \ref{fig:fraction_5-6Mpc}). (see Method Section for a detailed discussion)

\begin{figure*}
\centering
{
\includegraphics[width=\textwidth]{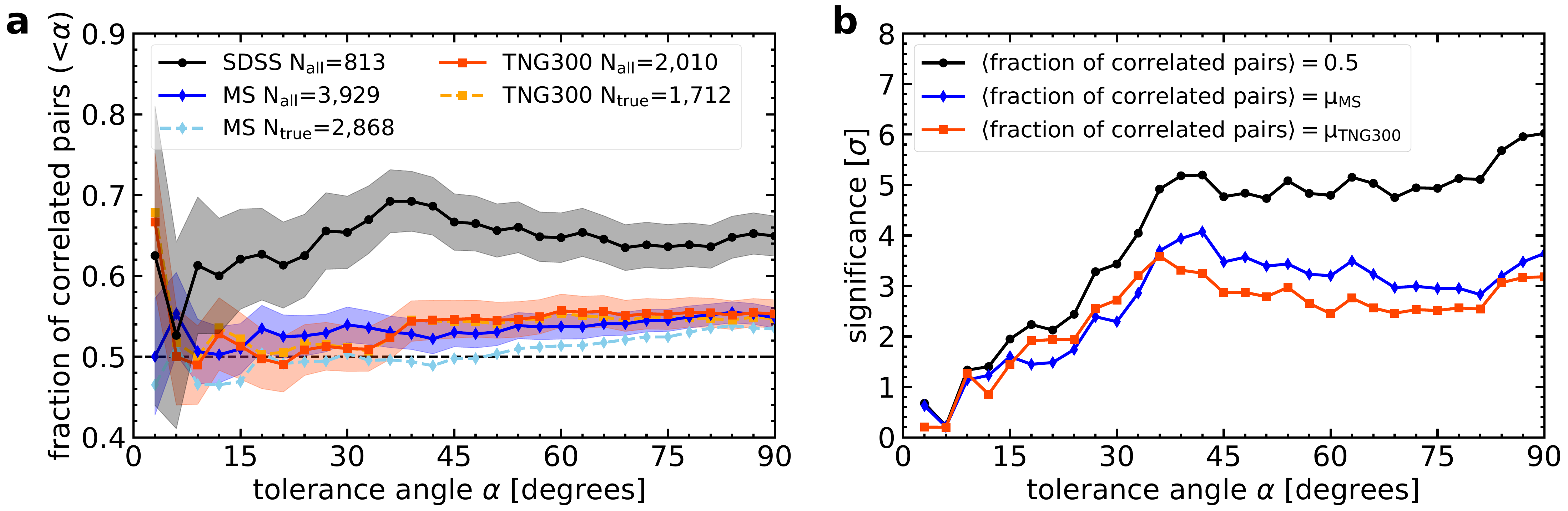}}

\caption{
    {\bf Contrasting satellite velocities in observations and cosmological simulations.} 
    \panel{a)} Comparison of the cumulative fraction of correlated pairs in SDSS and in the MS and TNG300 simulations as a function of tolerance angle. The blue solid curve with diamonds and orange red solid curve with squares show results in the MS and TNG300, respectively. Shadows with corresponding colours show their sample variances. Sky blue dashed curve with diamonds and orange dashed curve with squares show results for systems with `true' central galaxies in the MS and in TNG300 simulations. It shows that the results are not affected by the misidentification of primary galaxies significantly. The black solid curve with dots is the duplication of results in SDSS, whose fraction of correlated pairs is much higher. 
    \panel{b)} The significance of the excess of correlated pairs in SDSS compared to the MS (blue diamonds), to the TNG300 (orange red squares), and to a uniform distribution (black dots). 
    }
\label{fig:iso1_top2}
\end{figure*}

To fully account for the contamination from foreground and background galaxies, we apply the same sample selection and analysis to a full-sky mock semi-analytic galaxy catalogue based on the scaled Millennium simulation (MS)\cite{Henriques2015} and on a mock catalogue based on IllustrisTNG300-1 hydrodynamical simulation (hereafter TNG300) \cite{Nelson2018,Nelson2019}. Both adopted cosmological parameters consistent with the Planck\cite{Planck2014,Planck2016} measurements. For TNG300, we create eight mock catalogues by placing the observer in each of the eight corners of the simulation box at $z = 0$. In total, we find 3,929 and 2,010 galaxy systems that fulfil our selection criteria in the MS and TNG300, respectively. These systems have a similar redshift distribution (Extended Data Fig.\ \ref{fig:redshift}) and galaxy properties as the SDSS samples (Extended Data Fig.\ \ref{fig:compare}). As in SDSS, we find an excess of correlated pairs in both simulations, but the correlated pair fractions at $\alpha=90^{\degree}$ are lower, 54.8\% in the MS, and 55.3\% in TNG300 (Fig.\ \ref{fig:iso1_top2}a). For around 27.0\% of our MS sample, the selection procedure misidentifies a satellite galaxy as the primary galaxy. To test the impact of this effect, we calculate the correlated pair fraction using only systems around ‘true’ central galaxies (i.e. remaining 73.0\% of the sample), which also present an excess of the fraction of correlated pairs but the fraction at $\alpha=90^{\degree}$ decreases to 53.4\% in MS
and 54.7\% in the TNG300. The misidentification of primary galaxies does not change our conclusions.

The SDSS fraction of correlated pairs is significantly higher than the MS and the TNG300 results at $\alpha > 6^{\degree}$. In Fig.\ \ref{fig:iso1_top2}b we quantify the statistical robustness of the discrepancy. The excess in observations compared to simulations attains a maximum value of 4.1$\sigma$ ($p$-value of 2.1 $\times10^{-5}$) and 3.6$\sigma$ ($p$-value of 1.6 $\times10^{-4}$) with respect to the MS and the TNG300 predictions at $\alpha = 42^{\degree}$ and $\alpha =36^{\degree}$, respectively. This provides strong evidence of a deviation between observations and the predictions of $\Lambda$CDM cosmological simulations.  
In the Methods section, we describe further tests, such as stricter isolation criteria when selecting the sample. It shows that the measured excess in observations and its discrepancy with theoretical predictions are highly robust (Extended Data Fig.\ \ref{fig:iso2_top2}). 

It is possible that the excess of correlated satellite pairs results from recently accreted/infalling satellites. The most massive satellites fall along the dominant filament\cite{Shao2018} and are predicted to have correlated velocities at accretion. Such filaments have typical sizes larger than the 1 Mpc radius of our primaries\cite{Cautun2014} and can easily inject satellites on opposite sides of the central galaxy, which would lead to an excess of pairs with counter-rotating velocities. Since the orbital periods vary between satellites, this excess is probably short-lived. If we split the SDSS satellite sample according to the projected distance from the centre of the primary, we find that pairs of satellites with larger projected distances to their primaries ($r_{\rm p}>$ 642 Kpc, which is the median satellite distance) have a higher correlated fraction at $\alpha = 90^{\degree}$ than pairs of satellites closer to their primaries ($r_{\rm p}<$ 642 Kpc) (69.1\% versus 61.8\%). This provides indirect evidence for our hypothesis since recently accreted/infalling satellites are more likely to be found at higher radial distances. This hypothesis can also explain the difference between our excess of correlated pairs and the (still debated\cite{Cautun2015,Phillips2015}) slight excess of anti-correlated pairs found for galactic-mass hosts\cite{Ibata2014N}. Higher mass systems, such as our sample, are younger and formed later than galaxies like the Milky Way, and the correlated excess could be a transitory feature most pronounced when the host dark matter halo has just assembled most of its mass.  

To explain the discrepancy between the data and cosmological simulations, we further suggest that the observed systems are younger than those predicted by the $\Lambda$CDM model with Planck cosmological parameters (see Extended Data Fig.\ \ref{fig:fraction_z1}). Alternatively, galaxy sample selections rely on galaxy properties, which could vary if the sub-grid physics are changed. For example, reducing the stellar mass of recently accreted satellites/infalling galaxies in the model could result in a larger selection of interlopers, which, in turn, would reduce the excess of the fraction of correlated pairs. This study could thus pose further constraints on the sub-grid physics in cosmological simulations.

\clearpage

\pagestyle{empty}
\renewcommand{\thesubsection}{\Alph{subsection}}
\renewcommand{\thefigure}{\thesubsection .\arabic{figure}}
\renewcommand{\thetable}{\thesubsection .\arabic{table}}
\setcounter{figure}{0}
\setcounter{table}{0}

\begin{methods}
{\bf SDSS and MPA-JHU catalogue}\\
The Seventh Data Release of the Sloan Digital Sky Survey  has 11,663 deg$^2$ of imaging data and contains five-band ($ugriz$) photometry for 357 million distinct objects. There are over 1.6 million spectra in total, including 930,000 galaxies, over 9380 deg$^2$, centred at redshift $z \sim$ 0.1. The photometric data are complete down to $r$-band apparent magnitude of $\sim21$, and the extinction-corrected spectroscopy data are complete at $r\sim$17.72. We require all galaxies to have secure spectroscopic redshifts. Specifically, we select galaxies from the PhotoAllSpectra table.  We also account for the survey boundaries to ensure that most of the companions of our isolated primaries fall within the SDSS footprint. For this, we use the spherical polygons provided by the NYU-VAGC website to quantify the survey boundaries and the masked areas around bright stars. We remove any primaries centred on which the projected sky between 0.1 and 1 Mpc in radius has more than 5\% of the area lying outside the footprint covered by the VAGC spherical polygons. About 11 per cent of primaries are removed through this latter procedure, leading to our final sample of 813 galaxy systems. 

To have a complete sample, we use galaxies at low redshift, with $z < 0.065$. This is motivated two-fold. First, having a low redshift cut allows for a higher completeness sample down to lower stellar masses and thus a higher chance to find primary galaxies with two or more satellites. Secondly, the side-length of the TNG300 simulation box is ${\sim}\ 300$ Mpc, which is slightly larger than the comoving distance to redshift $z=0.065$. Thus, by using only galaxies with $z\ <\ 0.065$ we can build mock catalogues that are completely encapsulated within the TNG300 simulation box.
Satellite galaxies are ranked by their stellar masses, for which we use the stellar masses measured through the Spectral Energy Distribution (SED) fitting to SDSS $ugriz$ photometry \cite{Salim2007} from the MPA-JHU catalogue, rather than using spectral features \cite{Kauffmann2003}. As shown in Extended Data Fig.\ \ref{fig:galaxy_selection}, only those with stellar masses more massive than 10$^{10.08}$ $\rm M_{\odot}$ are complete at $z\ <\ 0.065$. 

\textbf{Toy models}\\
Here we investigate if the excess of satellite pairs with correlated velocities could be due to either central galaxies moving with respect to their host haloes or due to misidentifying satellite galaxies as centrals. To do so, we build two simplified models. In the first model, we test the effect of central galaxies having a non-zero velocity dispersion with respect to their dark matter haloes. We do so by generating line-of-sight velocities relative to the group centre where the satellites and centrals are sampled from two independent Gaussian distributions with dispersion $\sigma_s$ and $\sigma_c$, respectively. Then, if $\sigma_c=0$ the fraction of correlated satellite pairs is exactly 1/2, while if $\sigma_c=\sigma_s$ the correlated fraction is 2/3. Thus, for $0\leq\sigma_c\leq\sigma_s$ we have the fraction of correlated pairs that is between 50\% and 66.7\%.  In practice, we generate line-of-sight velocities for the satellites and centrals from three random Gaussian distributions. The positions of satellites  do not matter, and for simplicity we take the satellites to be on either side of the central (that is $\alpha = 0^\circ$). The velocities assigned to the centrals come from a Gaussian with width $\sigma_c = f_{\sigma} \sigma_s$, with $f_{\sigma}$ a fractional value between 0 and 1. Then we calculate the fraction of satellite pairs that have line-of-sight velocity offsets from the central of the same sign as a function of $f_{\sigma}$. The result of this simplified model is shown in Extended Data Fig.\ \ref{fig:toy-model}a and it shows that to obtain a 64\% excess of pairs with correlated velocities we need the central galaxies to have an average velocity dispersion $\sigma_c = 0.86\sigma_s$. Hydrodynamical simulations predict\cite{Ye2017} $f_{\sigma}$ = 0.27 for halos of $10^{13}\ \rm M_\odot$, the typical halo mass of our sample. This corresponds to the fraction of correlated pairs of 52\%, far below the observed value. It thus is highly unlikely to explain the observed excess of satellite pairs with correlated velocities.

In a second model, we study the effect of misidentifying the central galaxy. We assume that a random fraction $f_c$ of centrals are actually misidentified satellites, which could arise from the fact that a satellite galaxy has a higher stellar mass than its central. For the cases with misidentified centrals, the actual central galaxy is taken to be one of the satellites since is unlikely to have a central galaxy with stellar mass lower than that of the three most massive satellites. For simplicity, we assume that central galaxies are at rest with respect to their halos. In practice, we generate random Gaussian velocities for the two satellites and assign a velocity of 0 to the central. Then, with a probability $f_c$, we switch the velocity values of `satellite pairs' between the central and one of the satellites,  which is equivalent to misidentifying a central galaxy. Again, we calculate the fraction of opposed pairs of satellites that have line-of-sight velocity offsets from the central of the same sign as a function of $f_c$. The result is shown in Extended Data Fig.\ \ref{fig:toy-model}b. We find that the higher the chance to misidentify the central the higher the excess of pairs with correlated velocity, with a maximum value of 0.75 when all centrals are misidentified. To explain the observed excess, we need $f_c=0.56$, which is considerably higher than the 0.27 and 0.15 values we have measured in the MS and TNG300 simulations analyzed in this paper. Previous work\cite{Skibba2011} finds in SDSS that $f_c$ = 0.2 $\sim$ 0.4, corresponding to $\sim$ 55\% to 60\% pairs with correlated velocity. It is also not enough to explain the observed excess in the correlated satellite pairs.

Extended Data Fig.\ \ref{fig:toy-model}c shows the results obtained when accounting for both effects. Within a
possible range $f_\sigma \sim $ 0.27 and $f_c$ = 0.2 $\sim$ 0.4 as previously reported in the literature, the fraction of correlated pairs is
expected to be roughly 0.59. This fraction is lower than the 0.64 found in our study.

We then investigate the effect on the excess of the satellite pairs with correlated velocities of the foreground and background galaxies in Extended Data Fig.\ \ref{fig:fraction_5-6Mpc}. In practice, it examines the fraction of correlated pairs of neighbor galaxies within a projected shell between 5 Mpc and 6 Mpc and along a line-of-sight velocity difference of up to 1000 $\rm km\ s^{-1}$, centered on the primary galaxy. There are 660 systems that have at least two neighbours in the SDSS. In Extended Data Fig.\ \ref{fig:fraction_5-6Mpc}, the black solid curve with dots is the result of satellites located between 0.1 and 1 Mpc in projection while the cyan solid curve with squares shows the result of neighbours located between 5 and 6 Mpc in projection. The errors associated with the results are represented by the shadow regions, which denote the 1000 bootstrap errors. The 5-6 Mpc neighbours are observed to display a distribution close to random distribution. This indicates that the presence of foreground or background interlopers does not contribute to the excess of correlated pairs. 

{\bf Millennium simulation and mock catalogue}\\
The Millennium Simulation (MS) is a large N-body simulation of the standard $\Lambda$CDM cosmology. It is carried in a periodic cube of 500 $h^{-1}$Mpc side length and follows the evolution of 2160$^{3}$ DM particles from redshift 127 to the present day. The original cosmological parameters are consistent with the first-year Wilkinson Microwave Anisotropy Probe (WMAP) results: $\Omega_{\rm m}=0.25,\ \Omega_\Lambda= 0.75,\ \Omega_{\rm b}=0.045,\ h=0.73,\ \sigma_8=0.9$, and $n_{\rm s}=1$. It is then scaled to the Planck cosmology \cite{Planck2014} ( $\Omega_{\rm m} = 0.315,\ \Omega_\Lambda= 0.685,\ \Omega_{\rm b} = 0.049,\ h = 0.673,\ \sigma_8= 0.826,\ n_{\rm s} = 0.961$) using the Angulo\&White scaling technique \cite{Angulo2010,Angulo2015}. The rescaled MS corresponds to a simulation following 2160$^3$ particles in a periodic cube of 480.279 $h^{-1}$Mpc on a side, with each DM particle having a mass of 9.6 $\times$ 10$^{8}$ $h^{-1}\rm M_{\odot}$. All-sky light cones are constructed using galaxies catalogues which are generated by implementing the semi-analytic galaxy formation models onto the rescaled MS \cite{Henriques2015} dark matter trees. The mock is generated based on  the algorithm of the Mock Map Facility (MoMaF) \cite{Blaizot2005}.
We apply the same isolation criteria and satellite definition as for the observational data. This results in 3,929 systems with at least two satellite galaxies. The redshift distributions of the primary galaxies selected in SDSS and MS are similar to each other as shown in Extended Data Fig.\ \ref{fig:redshift}.

In MS, there are about 24\% of systems with at least one orphan satellite galaxy or a primary galaxy that has lost its subhalo either due to the stripping processes in the dense environment or due to the numerical effects. The positions of orphan galaxies are then calculated assuming dynamical friction-induced evolution. This effect has a very limited impact on our results. After removing systems with orphan galaxies, the average cumulative fraction of pairs with correlated velocities increases by 0.8\% for the full sample of the MS.

{\bf TNG simulations and mock catalogue}\\ 
The IllustrisTNG project consists of a suite of cosmological simulations that follow the formation and evolution of galaxies from $z = 127$ to the present day in cosmologically representative volumes. It adopts the moving mesh code AREPO\cite{Springel2010}, 
%\MCc{Cite Arepo paper!}
which solves the coupled equations of ideal magnetohydrodynamics and self-gravity, and cosmological parameters consistent with the  Planck results \cite{Planck2016} ($\Omega_{\rm m} = 0.3089,\ \Omega_\Lambda= 0.6911,\ \Omega_{\rm b} = 0.0486,\ h = 0.6774,\ \sigma_8= 0.8159,\ n_{\rm s} = 0.9667$). These simulations include prescriptions of gas cooling, star formation, stellar evolution, chemical enrichment, stellar feedback, supermassive black hole formation, AGN feedback, etc. It has been proven successful in reproducing many observational results both in the Local Universe and at high redshifts \cite{Nelson2018,Pillepich2018, Springel2018}. To have better statistics, we use TNG300, which was performed in a box of side 302.6~Mpc. It follows the evolution of $2500^3$ DM particles of mass $5.9\ \times\ 10^7\ \rm M_{\odot}$, and $2500^3$ gas cells of mass 1.1 $\times$ 10$^7$ $\rm M_{\odot}$. Since the stellar mass is underestimated in TNG300, we multiply the stellar mass of each galaxy by 1.4 as suggested in their main paper \cite{Pillepich2018}, which results in a galaxy stellar mass function that is in good agreement with observations. The mock galaxy catalogue was generated by placing the observer at each of the eight corners of the simulation box at $z = 0$. We assign each galaxy a redshift based on its line-of-sight distance and peculiar velocity. We apply the same isolation criteria and satellite definition as for the observational data. This results in 2,010 unique systems with at least two satellite galaxies. The redshift distributions of the primary galaxies selected in SDSS and TNG300 are similar to each other (Extended Data Fig.\ \ref{fig:redshift}).

To measure the completeness value, we use the TNG100 simulation which has a box size of 110.7 $\rm Mpc^3$. It follows the evolution of $1820^3$ DM particles of mass 7.5 $\times\ 10^6 \rm\ M_{\odot}$ and $1820^3$ gas cells of mass 1.4 $\times\ 10^6 \rm\ M_{\odot}$. We divide redshifts from 0 to 0.12 to 24 bins with bin size being 0.005. For each redshift, we put the galaxies in the simulated box at the corresponding luminosity distance and select the continuous stellar mass bins in which there are more than 90\% galaxies with $r$-band apparent magnitude smaller than 17.72. We define the lowest stellar mass bin as the complete stellar mass. For $z\leq0.05$, we use the box of snapshot 99 ($z=0$), while we use the box of snapshot 91 ($z=0.1$) for $0.05<z\leq0.12$. The black solid line in Extended Data Fig.\ \ref{fig:galaxy_selection} shows the 90\% completeness value as a function of redshift. 

In Extended Data Fig.\ \ref{fig:compare}, we present comparisons between observed and simulated systems, including the angular distributions between the pairs, the relative line-of-sight velocity distributions, the stellar mass difference distributions between satellite and primary galaxy, the $r$-band absolute magnitude difference distributions, the stellar mass distributions of primary and satellite galaxies. Extended Data Fig.\ \ref{fig:compare}a and Extended Data Fig.\ \ref{fig:compare}b show that the angular distribution between the pairs and relative line-of-sight velocity distributions in simulations closely resemble those seen in observations. Extended Data Fig.\ \ref{fig:compare}c and Extended Data Fig.\ \ref{fig:compare}d reveals typically more than a factor of 3 difference in stellar mass and a more than 1 mag difference in $r$-band magnitude. The stellar mass difference in SDSS and MS is similar, albeit somehow smaller, than that in TNG300. Extended Data Fig.\ \ref{fig:compare}e and Extended Data Fig.\ \ref{fig:compare}f show that the stellar mass distributions in MS and SDSS are very similar both for primaries and satellites. In TNG300, the selected primaries are
somehow more massive and satellites are slightly less massive. Overall, we observe a good agreement between the simulated and observed distributions. 

\textbf{{Stricter isolation criteria}}\\ 
To further investigate the robustness of our results, we redo the analysis by using the more strict isolation criteria. We test the results by requiring each primary to be the brightest galaxy within a larger volume, $r_{\rm p}<$ 2 Mpc and $|\Delta v|$ $<$ 2000 km s$^{-1}$. This leads to a $\sim$21\% reduction in sample size. One of the advantages of the stricter isolation criteria is an increase of the sample purity (i.e. the fraction of primaries which are actual central galaxies) which now is 78.0\% in MS and 91.4\% in TNG300. While we cannot calculate the purity for SDSS, we would expect a similarly high purity too. The SDSS fraction of correlated pairs is nearly the same as in our previous results, with a value of 63.4\% at 90$^{\degree}$ (Extended Data Fig.\ \ref{fig:iso2_top2}a).
Compared to the null expectation of no correlated satellite velocities, the SDSS result represents a $4.5\sigma$ excess ($p$-value $=3.4\times10^{-6}$) at 90$^{\degree}$ (Extended Data Fig.\ \ref{fig:iso2_top2}b). Similarly, the SDSS correlated pair fraction is higher than the simulations' predictions with a significance whose maximum is $3.4\sigma$. The smaller sample size is the main reason why these significance values are lower than the ones found for the less restricted isolation conditions.

We also test the results by requiring a larger difference in magnitude between primary and satellite galaxies within a projected separation of 0.5 Mpc. More specifically, a minimum difference of 1.5 and 2 magnitudes is chosen instead of the 1 magnitude adopted for our primary results. We compare the fraction of correlated pairs among samples with minimum differences of 1, 1.5, and 2 magnitudes between their primary and satellite galaxies. These three samples show a similar excess in the fraction of correlated pairs, suggesting that more stringent conditions on the magnitude differences
between primaries and satellites do not affect our conclusion. (Extended Data Fig.\ \ref{fig:iso2_top2} cdef)

\textbf{Comparing excess of correlated satellite galaxies at different redshifts}\\
The excess of correlated satellite galaxy pairs with diametrically opposite positions around massive galaxies in SDSS suggests that the Universe is younger than what is predicted by the Planck cosmology. We use simulations to examine this hypothesis by comparing the excess of correlated satellite galaxies observed in the local universe to their counterparts at higher redshifts, when the universe was younger. We analyze the snapshots corresponding to $z=1$ in both MS and TNG300 and conduct a similar analysis. As shown in Extended Data Fig.\ \ref{fig:fraction_z1}, we find that in both MS and TNG300, the fractions of correlated pairs of satellite galaxies are significantly larger at $z=1$ compared to their predicted values at $z=0$. This provides support for our explanation that the observed systems may indeed be younger than what is expected based on the $\Lambda$CDM model with the Planck cosmological parameters.

\end{methods}

\begin{addendum}
 \item[Data availability]
 All data used in this work is publicly available. The SDSS data are available at the Sloan Digital Sky Survey (https://www.sdss.org/). The data for TNG300 and TNG100 can be accessed at https://www.tng-project.org. The data for Millennium can be accessed at http://gavo.mpa-garching.mpg.de\\
 /Millennium/. The other data that support the results of this study are available from the corresponding author upon reasonable request.
\item[Code availability]
We use standard data reduction tools in Python environments.
 \item[Acknowledgements] 
   We thank Prof. Dr. Simon White for his constructive comments. This work is supported by the National SKA Program of China (No. 2022SKA0110201 and 2022SKA0110200), the National Key Research and Development of China (grant number 2018YFA0404503), the National Natural Science Foundation of China (NSFC; 12033008, 12273053, 11988101 and 12022307), the K.C. Wong Education Foundation, CAS Project for Young Scientists in Basic Research Grant (No. YSBR-062) and the science research grants from China Manned Space Project with No.CMS-CSST-2021-A03 and NO. CMS-CSST-2021-B03. M.C. acknowledges the support from the EU Horizon 2020 research and innovation programme under a Marie Sk{\l}odowska-Curie grant agreement 794474 (DancingGalaxies). W.W. acknowledges the support from the Yangyang Development Fund. Q. Guo thanks  European Union’s HORIZON-MSCA-2021-SE-01 Research and Innovation programme under the Marie Sk{\l}odowska-Curie grant agreement number 101086388.
 \item[Author contributions] 
 Qi Guo led and played a part in all aspects of the analysis. Qing Gu compiled the data and carried out most of the data reduction and analysis, and wrote the manuscript. All authors contributed to the analysis, and to the writing of the manuscript.
 \item[Competing financial interests]
 The authors declare no competing interests. 
 \item[Additional information] 
 Supplementary information is available in the on-line version of the paper.
 
 Correspondence and requests for materials should be addressed to Qi Guo (guoqi@nao.cas.cn) and Shi Shao (shaoshi@bao.ac.cn).
 \end{addendum}

%% Here is the endmatter stuff: Supplementary Info, etc.
%% Use \item's to separate, default label is "Acknowledgements"

%%
%% TABLES
%%
%% If there are any tables, put them here.
%%
\newpage
\appendix

\renewcommand{\thetable}{Table \arabic{table}} 
\renewcommand{\figurename}{Extended Data}
\renewcommand{\tablename}{Extended Data}
\captionsetup[figure]{name={Extended Data Fig.}}
\renewcommand{\thefigure}{\arabic{figure}} 

\begin{figure*}
\centering
\includegraphics[width=0.7\textwidth]{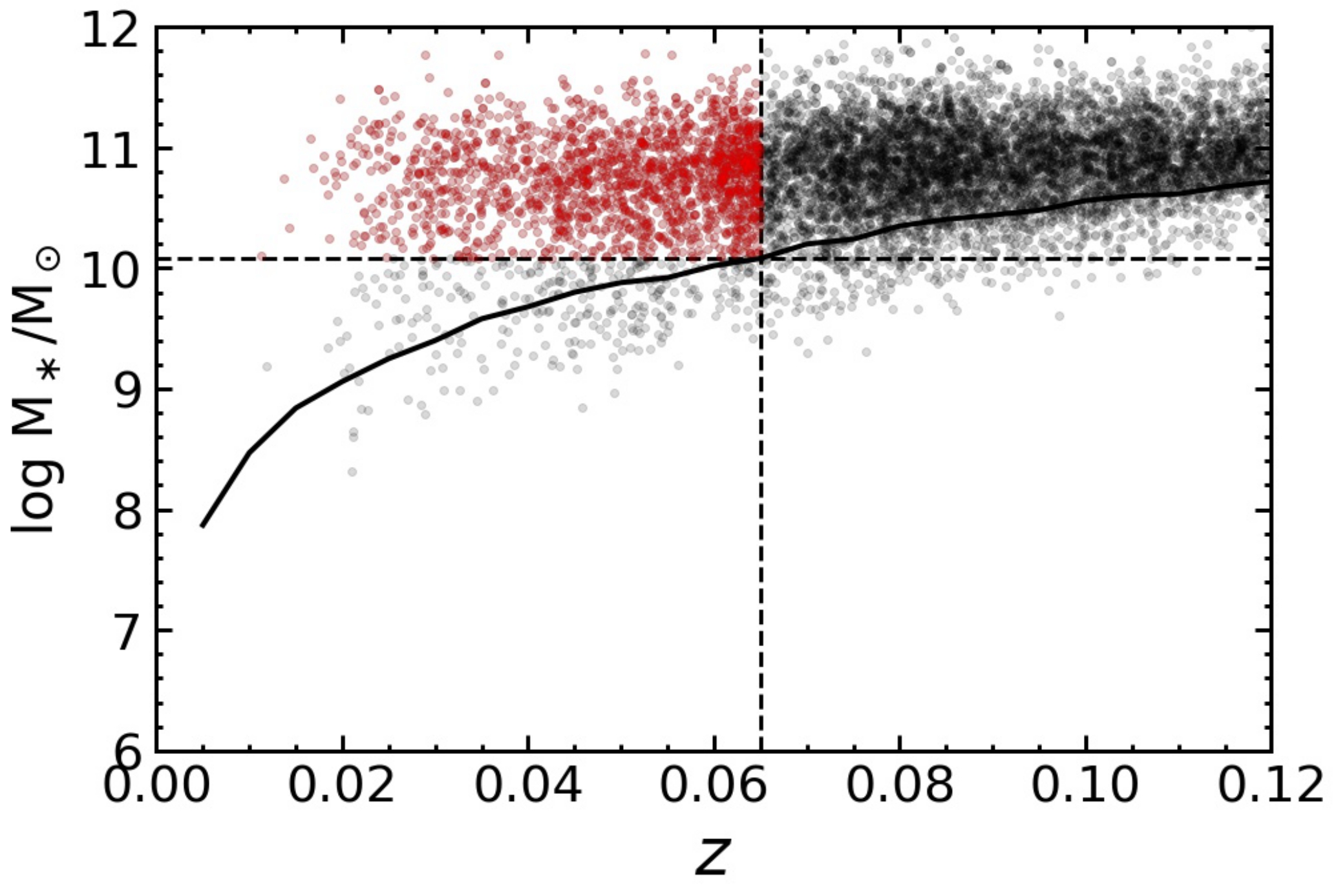}
    \caption{{\bf Galaxy selections.} Redshift versus stellar mass of the two most massive satellites in SDSS. The black curve shows the 90\% completeness stellar mass value as a function of redshift. The dashed vertical line denotes $z = 0.065$, which is the redshift cut of primaries and satellites, and the dashed horizontal line corresponds to galaxies more massive than 10$^{10.08}$ $\rm M_{\odot}$, which is the lower mass threshold we use to select satellite galaxies.} 
\label{fig:galaxy_selection}
\end{figure*}

\begin{figure*}
\centering
\includegraphics[width=\textwidth]{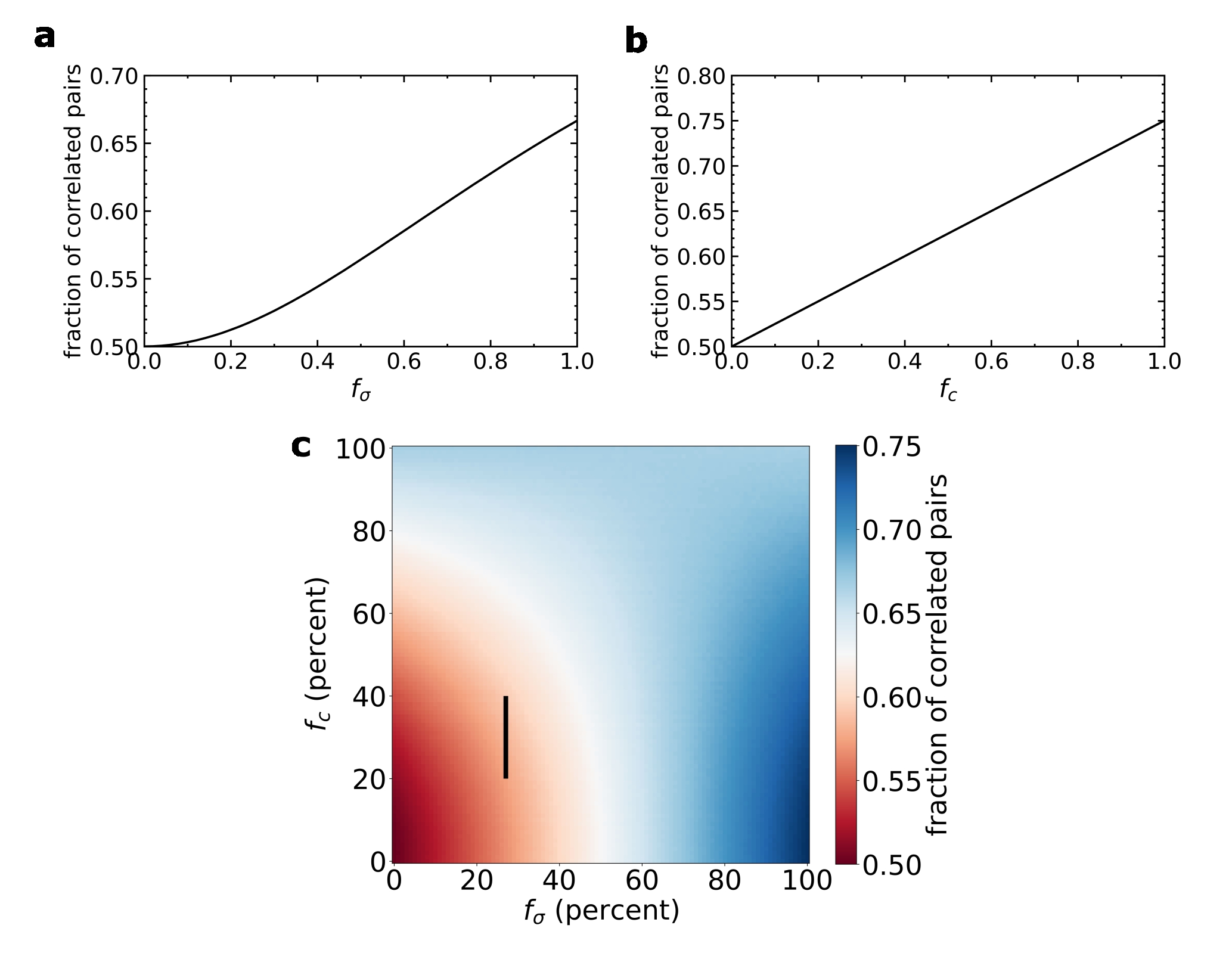}

    \caption{{\bf The excess of satellite pairs with correlated velocity in simple models.} \panel{a)} Central galaxies move with respect to the host dark matter halo with a velocity dispersion $\sigma_c=f_{\sigma} \sigma_s$, where $f_{\sigma}$ is a fractional value between 0 and 1. \panel{b)} A fraction $f_c$ of centrals that are actually misidentified satellite galaxies. \panel{c)} The fraction of correlated pairs considering both the effects of velocity dispersion of centrals relative to their
halos and the misidentification of centrals. The black solid line illustrates the corresponding region reported
in the literature.}
\label{fig:toy-model}
\end{figure*}

\begin{figure*}
\centering
\includegraphics[width=0.7\textwidth]{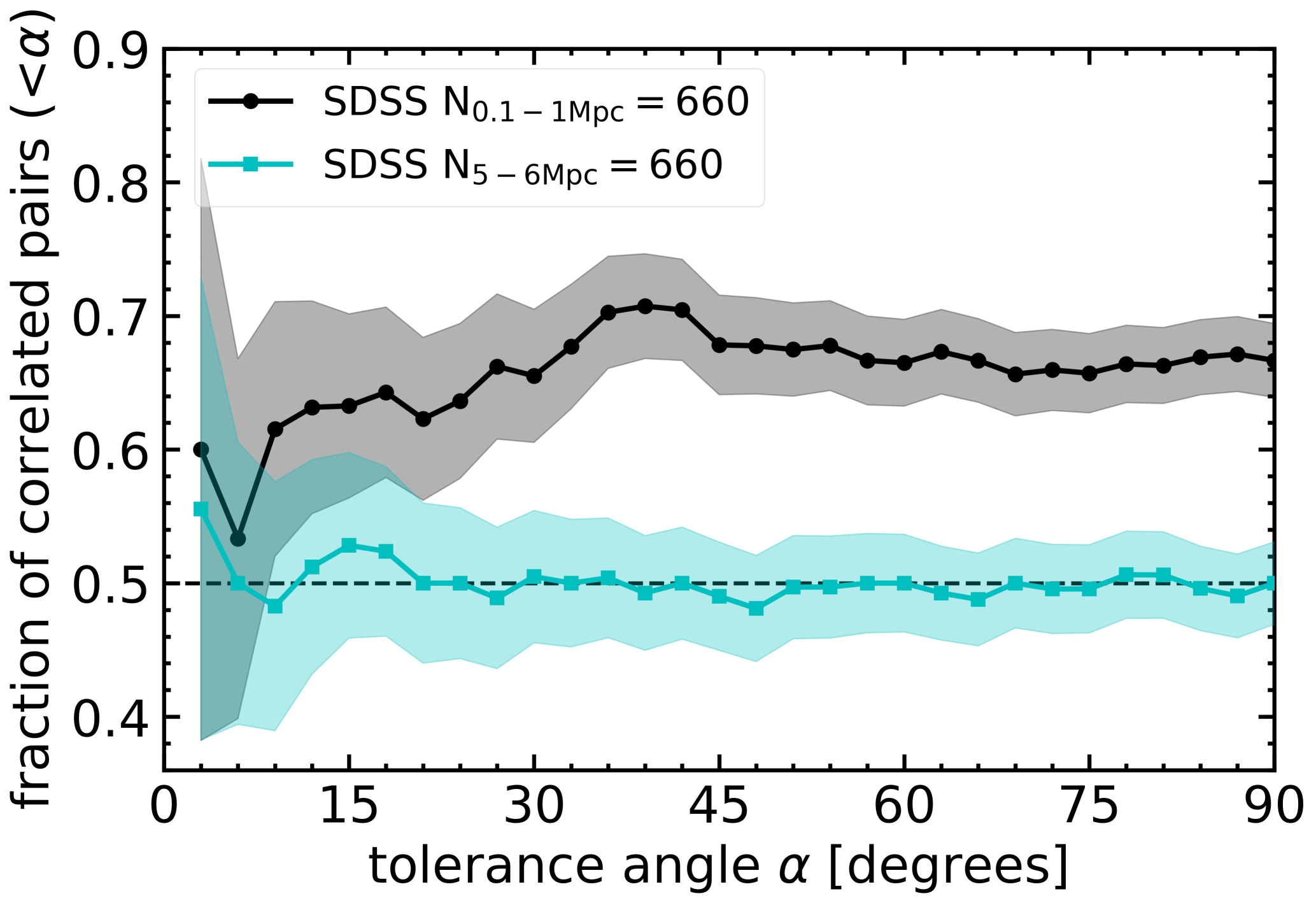}
    \caption{{\bf The velocity correlation of neighbour galaxy pairs within a more extended region in SDSS.} The comparison of correlated fraction between satellite pairs located from 0.1 Mpc to 1 Mpc in projection (black solid curve with dots) and neighbours located from 5 Mpc to 6 Mpc in projection (cyan solid curve with squares). The horizontal dashed line represents the random distribution. Shade regions represent the standard deviations of corresponding bootstrap samples.}     
\label{fig:fraction_5-6Mpc}
\end{figure*}

\begin{figure*}
\centering
\includegraphics[width=0.7\textwidth]{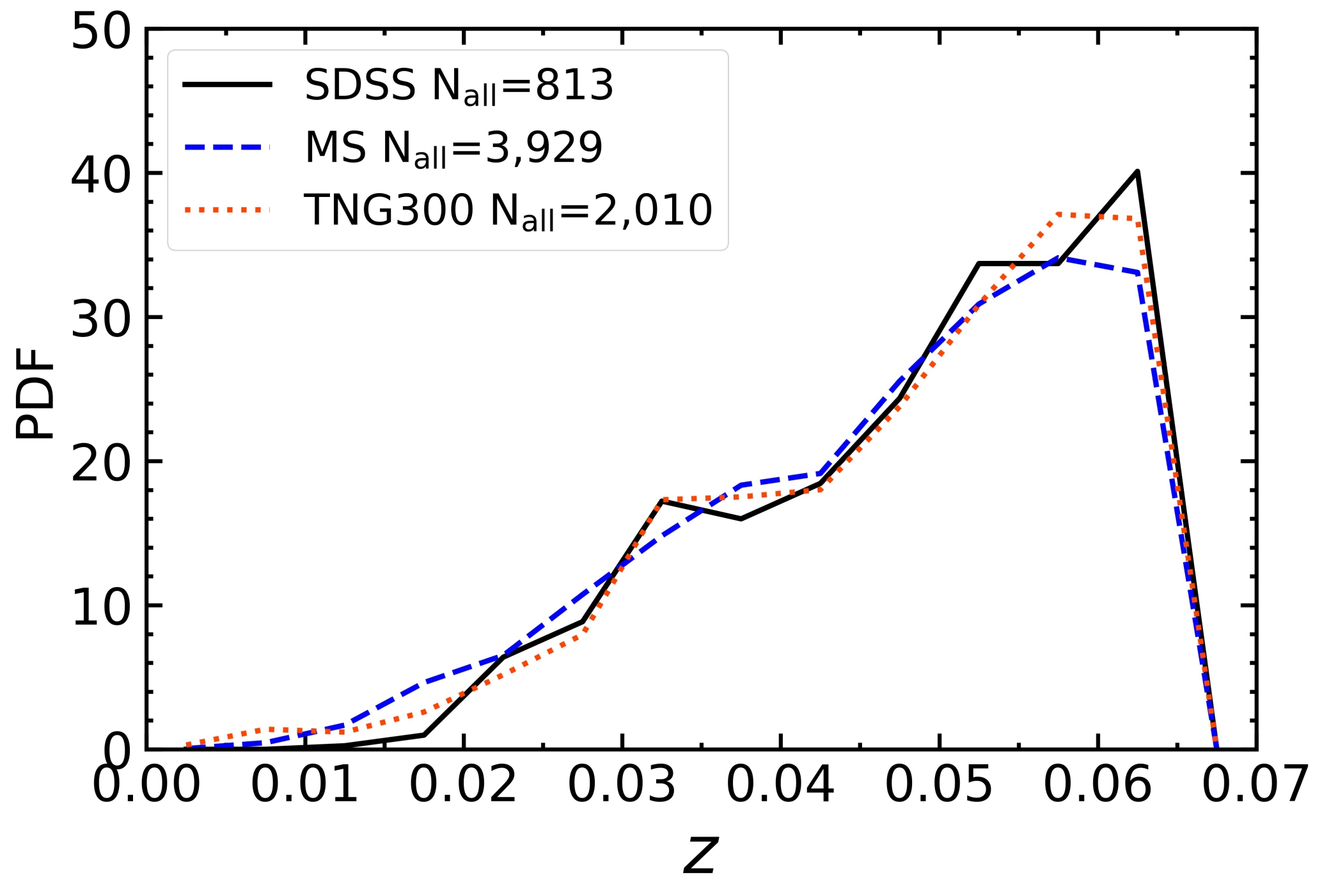}

    \caption{{\bf Redshift distributions.} Comparison of the redshift distributions of the primary galaxies in SDSS (black solid), MS (blue dashed) and TNG300 (orange red dotted).} 
\label{fig:redshift}
\end{figure*}

\begin{figure*}
\centering
\includegraphics[width=\textwidth]{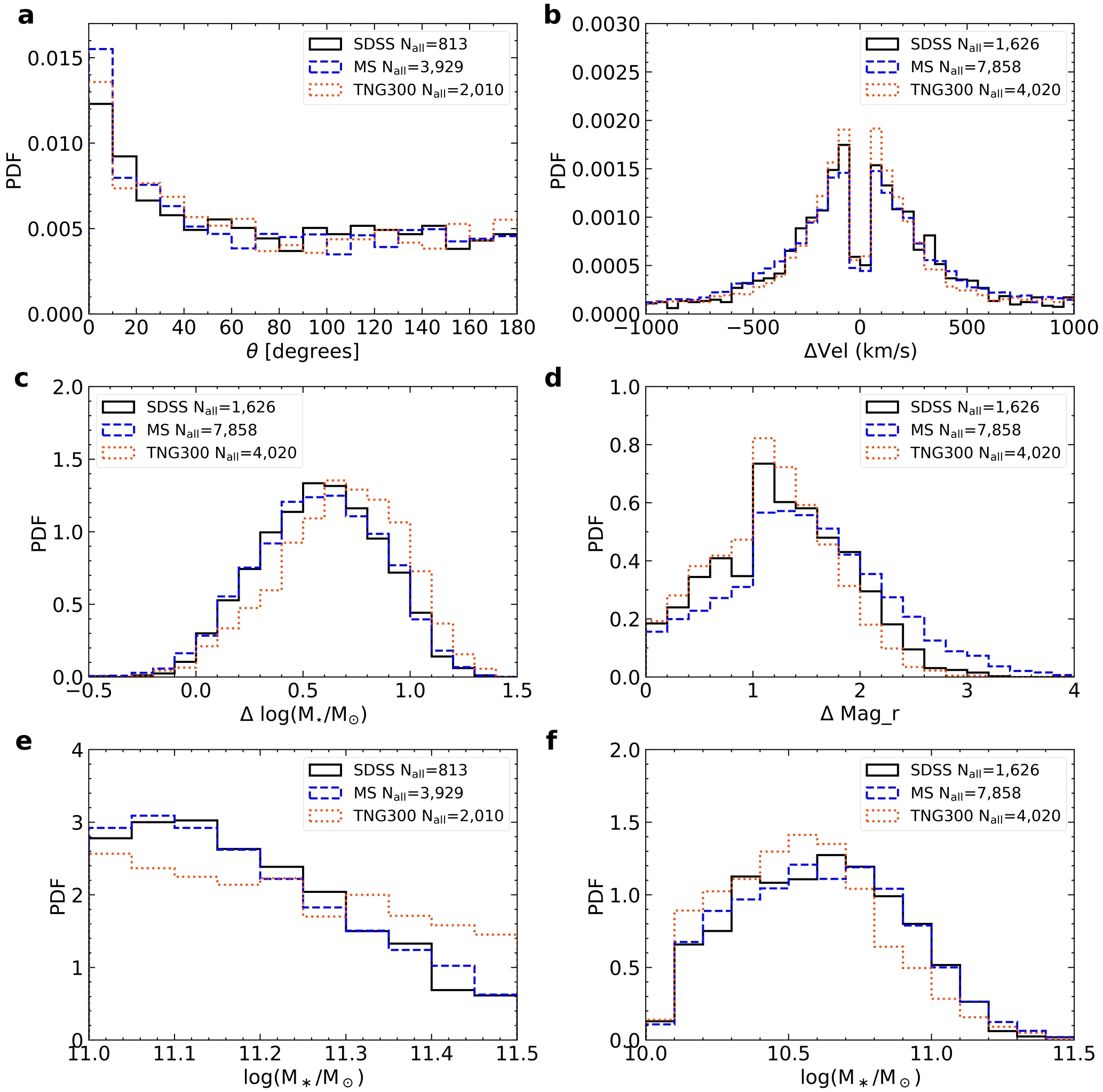}

    \caption{{\bf Comparison between observation and simulations.} \panel{a)} The angular distributions in SDSS and simulations. \panel{b)} The distributions of relative line-of-sight velocity difference in observation and simulations. \panel{c)} The distributions of the difference in stellar mass between satellite and primary in SDSS, MS and TNG300. \panel{d)} The distributions of the difference in absolute magnitudes between satellite and primary in SDSS, MS and TNG300. \panel{e)} The distributions of stellar mass of primary galaxies. \panel{f)} The distributions of stellar mass of satellite galaxies. Overall, we observe a good agreement between simulations and observations.} 
\label{fig:compare}
\end{figure*}

\begin{figure*}
\centering
\includegraphics[width=\textwidth]{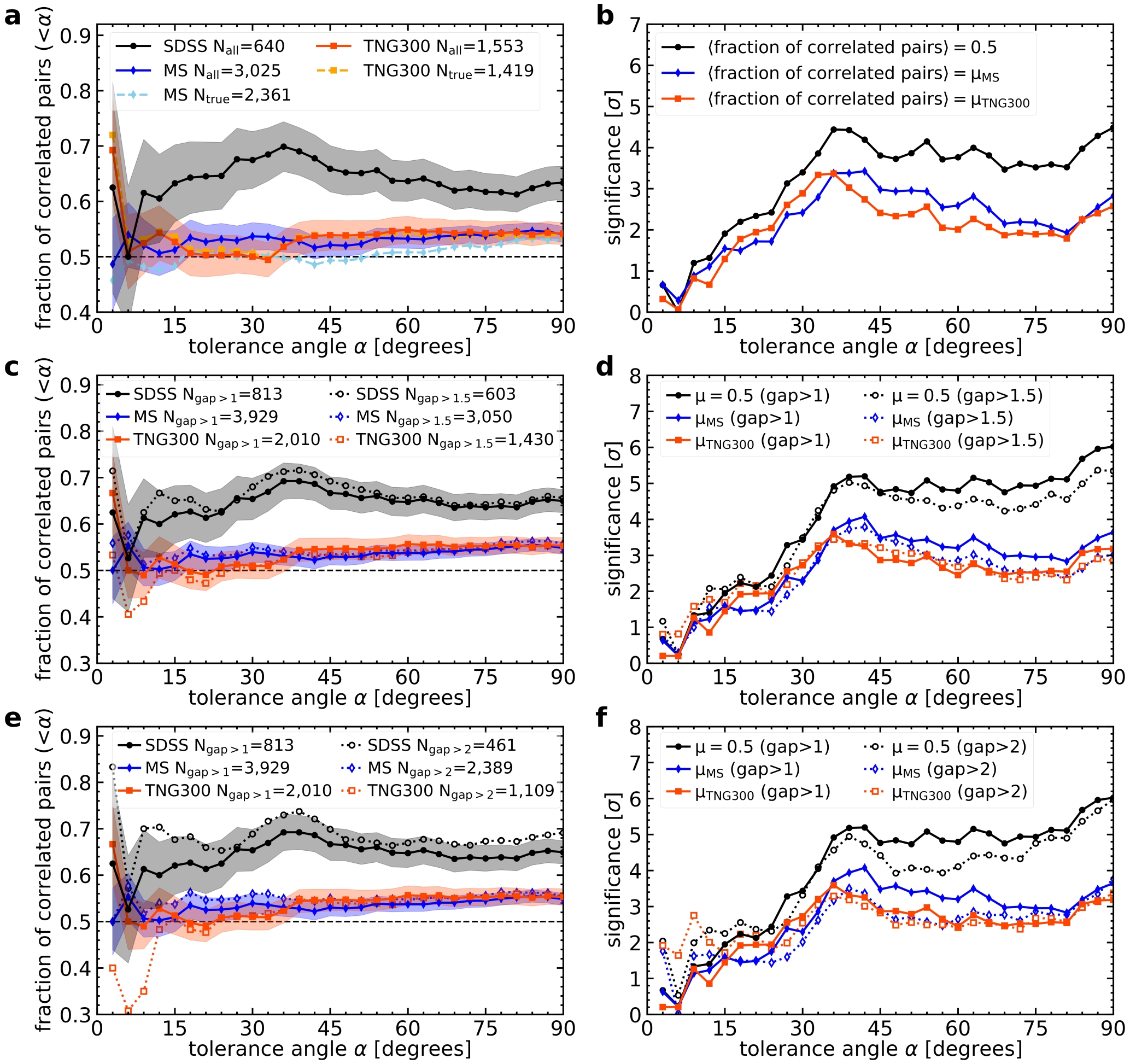}
    \caption{{\bf The velocity correlation of satellites for samples with stricter isolation criteria.} 
     \panel{a)} and \panel{b)} Same as Fig.\ \ref{fig:iso1_top2} in the main paper but for systems whose primaries are the brightest galaxies within a larger volume, $r_{\rm p}<$ 2 Mpc and $|\Delta v|$ $<$ 2000 km s$^{-1}$. It results in a $\sim$21\% reduction in sample size. In SDSS, the excess of satellite pairs with correlated velocities is similar to our main results. The significance of this excess is lower due to the smaller sample size. \panel{c)} and \panel{d)} Fraction of correlated pairs and the significance of the excess for systems which choose a larger minimum magnitude difference between primary and satellite galaxies within 0.5 Mpc projected distance, specifically 1.5 magnitudes. The results are compared to those obtained with a minimum magnitude difference of 1 magnitude, which is used in the main analysis. \panel{e)} and \panel{f)} Similar to \panel{c)} and \panel{d)} but for a minimum difference of 2 magnitude. The solid curves with solid markers and dashed curves with empty markers are for 1 and 1.5 (or 2) magnitude differences, respectively. Results from SDSS, MS and TNG300 are shown in black, blue and orange red curves, respectively. It demonstrates the robustness of our findings, even when implementing more stringent isolation criteria for the selection of primaries and satellites. The shaded regions in the left three panels correspond to the standard deviations of bootstrap samples.}
\label{fig:iso2_top2}
\end{figure*}

\begin{figure}
  \centering
  \hspace*{-0.5cm}
  \includegraphics[width=\textwidth,angle=0]{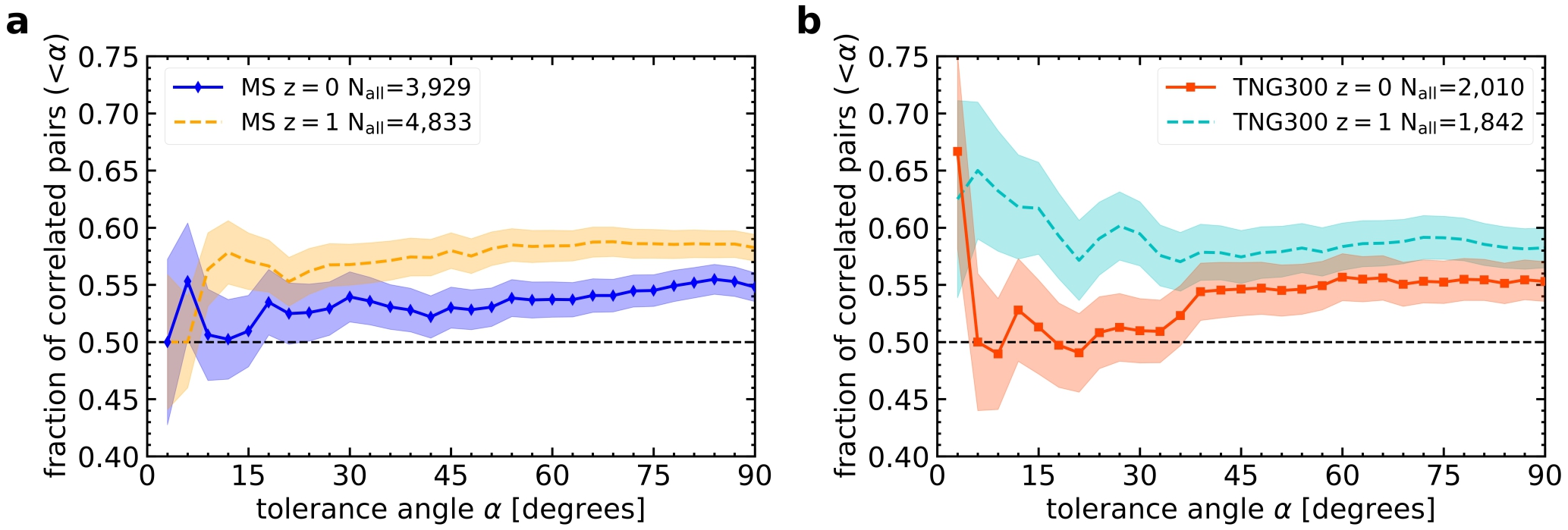}

  \caption{{\bf The velocity correlation of satellites at different redshifts in simulations.} \panel{a)} Cumulative fraction of correlated galaxy pairs as a function of tolerance angle in the MS. The blue solid line with diamonds represents the MS result at $z=0$, taken from Fig.\ \ref{fig:iso1_top2}a in the main text. The orange dashed line shows the cumulative fraction of correlated pairs at $z=1$. \panel{b)} Cumulative fraction of correlated galaxy pairs as a function of tolerance angle in the TNG300. The orange red solid line with squares duplicates the TNG300 result at $z=0$ from Fig.\ \ref{fig:iso1_top2}a in the main text. The cyan dashed line illustrates the cumulative fraction of correlated pairs at $z=1$. The shadow regions indicate the corresponding standard deviations of bootstrap samples.}
  \label{fig:fraction_z1}
\end{figure}

\end{document}